\documentclass[]{aa}
\usepackage{graphics}
\usepackage{amsmath}
\usepackage{epsfig}

\begin{document}

 \title{Do dusty A stars exhibit accretion signatures in their photospheres?}
 \titlerunning{accretion signatures in A stars}

 \author{Inga~Kamp \inst{1}, Marc Hempel \inst{2}, Hartmut Holweger \inst{3}}
 \authorrunning{Kamp et al.}
 \offprints{Inga~Kamp}
 \mail{kamp@strw.leidenuniv.nl}
 \institute{Leiden Observatory, Niels Bohrweg 2, PO Box 9513, 2300 RA Leiden, 
            The Netherlands \and Hamburger Sternwarte, Gojenbergsweg 112,
            21029 Hamburg, Germany \and Institut f\"{u}r Theoretische
            Physik und Astrophysik, Universit\"{a}t Kiel, 24098 Kiel, Germany}
 \date{Received 31 January 2002/ Accepted 29 March 2002}

 \abstract{We determined 
           abundances of O, Ca, Fe, Ba and Y for a sample of dusty and dust-free A stars, taken 
           from the list of Cheng et al. (\cite{Cheng92}). Five of the stars have an 
           infrared-excess due to circumstellar dust. Ongoing accretion 
           from their circumstellar surroundings might have modified the 
           abundances in the photospheres of these stars, but our results 
           clearly show, that there is {\sl no difference} in the
           photospheric composition of the dusty and dust-free stars.
           Instead all of them show the typical diffusion pattern which
           diminishes towards larger rotational velocities.
  }

 \maketitle

\section{Introduction}
IRAS observations have shown that about 20\% of all nearby A~stars are 
surrounded by dust (Cheng et al. \cite{Cheng92}); the most prominent
prototypes of these dusty A~stars are Vega and $\beta$~Pictoris. 
In order to understand the nature of these dusty A~stars, Holweger
\& Rentzsch-Holm (\cite {HRH95}) and Holweger et al. (\cite{HHK99}, 
in the following HHK99)
searched for gas in the circumstellar (CS) environment of a sample
of dusty and dust-free A~stars. They found narrow absorption
features in the \ion{Ca}{ii}\,K and the \ion{Na}{i}\,D lines in
about 30\% of the stars. The occurrence of these features shows 
a strong correlation with the projected rotational velocity, which 
suggests the presence of gas concentrated in a disk-like structure, 
and therefore proves that these imprints on the photospheric 
spectrum originate in the {\sl circum}stellar rather than in 
the {\sl inter}stellar medium (Holweger \& Rentzsch-Holm \cite{HRH95}). 

Since main-sequence A stars possess extremely shallow 
surface convection zones, their atmospheres are stable enough to allow 
{\sl diffusion} processes to occur (Michaud \cite{Mich70}). 
This leads to a variety of peculiarities like the AmFm-phenomenon, which 
has been specifically investigated in open clusters with well-known ages 
(see e.g. Alecian \cite{Ale96}, Hui-Bon-Hoa \& Alecian \cite{Hui98} and 
references therein). Moreover, the shallow convection zones allow any
contamination from circumstellar or interstellar matter to show
up in the photospheres of these stars. The metal underabundances of 
a small subgroup of A stars, such as the $\lambda$~Bootis stars,
can be explained in terms of {\sl accretion} of metal-deficient 
gas from their circumstellar environment (Venn \& Lambert \cite{VL90}). 
In this subgroup of stars the dust phase is enriched in condensable 
elements whereas the gas phase is heavily metal-depleted, except for 
those elements which have a low condensation temperature, like C, N, O, and S.
Conversely, in the framework of the ''planet migration`` model, the 
metal overabundances of solar-type stars with planets are 
attributed to the accretion of rocky planets onto the star (Smith et al.
\cite{Smith}). A recent study of non-interacting solar-type
binary stars revealed the existence of pairs of stars, where one component 
was more metal-rich than the other (Gratton et al. \cite{Gratton}).
There is a trend towards increasing abundance difference with
condensation temperature, pointing towards the accretion of 
dust-rich or rocky material onto the stellar surface from either 
the inner part of a protoplanetary disk or from rocky planets.
Despite the wealth of literature dicussing the circumstellar disks and
peculiar abundance patterns in main-sequence A stars, a definite 
connection between these phenomena has been discussed but never really
proven.

\begin{table*}[ht]
\caption{Our program stars. Visual magnitudes and spectral types are from the 
Bright Star Catalogue (Hoffleit \& Warren \cite{BSC}). Columns 6\,--9 indicate 
whether we have obtained spectra for the O, Na, Ca, and Ba lines. CS denotes whether 
the star shows narrow absorption components in \ion{Ca}{ii}\,K. The entry 'dusty' was 
adopted from the Cheng et al. (\cite{Cheng92}) compilation.}
\begin{tabular}{rlrrlccccccc} 
\hline
\rule{0mm}{6.0mm} 
HR~  &Name               &HD~~    &  V~~  &Spectral~Type& O  & Na & Ca & Ba &  CS? & dusty? & Remarks~\\[2mm]
\hline\\
 553 & $\beta$\,Ari      &   11636 & 2.64 & A5\,V       & +  & -  & +  & +  &   -  &    -   &  ~\\
 804 & $\gamma$\,Cet     &   16970 & 3.47 & A3\,V       & +  & -  & +  & +  &   -  &    -   &  ~\\
1483 &                   &   29573 & 5.01 & A2\,IV      & +  & -  & +  & +  &   -  &    +   &  ~\\
1666 & $\beta$\,Eri      &   33111 & 2.79 & A3\,III     & +  & -  & +  & +  &   -  &    -   &  ~\\
1989 & 131\,Tau          &   38545 & 5.72 & A3\,Vn      & +  & +  & +  & -  &   +  &    -   & $\lambda$~Boo ~\\
2491 & $\alpha$\,CMa     &   48915 &-1.46 & A1\,Vm      & +  & -  & +  & -  &   -  &    -   &  ~\\
2763 & $\lambda$\,Gem    &   56537 & 3.58 & A3\,V       & +  & -  & +  & +  &      &    -   &  ~\\
3083 &                   &   64491 & 6.23 & A3\,IVp     & +  & -  & +  & -  &      &    -   & binary (1) $\lambda$~Boo ~\\
3569 & $\iota$\,UMa      &   76644 & 3.14 & A7\,IV      & +  & -  & +  & +  &      &    -   &  ~\\   
4295 & $\beta$\,UMa      &   95418 & 2.37 & A1\,V       & +  & -  & +  & +  &   -  &    +   &  ~\\   
4534 & $\beta$\,Leo      &  102647 & 2.14 & A3\,V       & +  & +  & +  & +  &   -  &    +   &  ~\\   
4828 & $\rho$\,Vir       &  110411 & 4.88 & A0\,V       & +  & -  & -  & -  &   -  &    +   &  $\lambda$~Boo~\\   
5351 & $\lambda$\,Boo    &  125162 & 4.18 & A0p         & -  & -  & +  & +  &   -  &    +   &  $\lambda$~Boo~\\   
5531 & $\alpha^{2}$\,Lib &  130841 & 2.75 & A3\,IV      & +  & -  & +  & +  &   -  &    -   & binary (4)~\\   
5793 & $\alpha$\,CrB     &  139006 & 2.23 & A0\,V+G5\,V & +  & +  & +  & +  &      &    +   & binary (3)~\\   
5895 & 36\,Ser           &  141851 & 5.11 & A3\,Vnp     & -  & -  & +  & -  &      &    -   & $\lambda$~Boo ~\\   
6378 & $\eta$\,Oph       &  155125 & 2.43 & A2\,V       & +  & -  & +  & -  &   -  &    -   & binary (2) ~\\   
6556 & $\alpha$\,Oph     &  159561 & 2.08 & A5\,III     & +  & +  & +  & +  &   +  &    -   &  ~\\  
7001 & $\alpha$\,Lyr     &  172167 & 0.03 & A0\,Va      & +  & -  & +  & -  &   -  &    +   & $\lambda$~Boo ~\\   
8728 &$\alpha$\,PsA      &  216956 &  1.16 & A3\,V      & -  & -  & +  & -  &   -  &    +   &  ~\\
8947 & 15\,And           &  221756 & 5.59 & A1\,Vp      & -  & -  & +  & -  &   -  &    -   & $\lambda$~Boo ~\\   
\hline\\
\end{tabular}

1: Kamp et al. (2001); 2: Ten Brummelaar (2000); 3: SIMBAD; 4: BSC (Hoffleit \& Warren 1991)
\label{stars}
\end{table*}

A study on the surface composition of $\beta$~Pictoris carried out by 
Holweger et al. (\cite{Holweger97}) showed that this example of a star 
with CS matter, has solar abundances 'within a factor two or less' and 
therefore reveals no signs of accretion. On the other hand Vega, another
example of a star with CS matter, is often called a ''mild'' 
$\lambda$~Bootis star (Venn \& Lambert \cite{VL90}; Lemke \& Venn 
\cite{Lemke3}), because it shows moderate metal underabundances
that can be explained by the accretion of gas depleted in condensable 
elements. In general this raises some questions when applied to normal A~stars 
(i.e. stars without known peculiarities): Is the result obtained for 
$\beta$~Pic or Vega outstanding?: Can the presence of CS matter 
affect the abundances?: Is it possible to detect these signs of 
accretion? 

These questions can be answered by an extended search for the
differences between the abundance pattern of dusty and dust-free 
normal A~stars. Therefore, we have carried out a detailed abundance analysis 
of oxygen, calcium, barium, yttrium, and iron in 
a sample of nearby dusty and dust-free normal 
A stars. These elements are selected as representatives of the light 
elements with low condensation temperatures (O), and the heavy elements 
with higher condensation temperatures (Ca, Ba, Y, and Fe). Ba and Ca are 
strongly affected by diffusion whereas Fe is not. We scrutinized our
sample to detect systematic differences in surface composition between 
dusty and dust-free stars.

\section{The stellar sample}

We have 
compiled a sample of 21 northern sky object to compare the abundances 
of dusty and dust-free A~stars. It contains 
15 normal A~stars taken from Cheng et al. (\cite{Cheng92}), who
cross-correlated the catalogue of A stars within 25~pc 
(Woolley et al. \cite{W70}) with the IRAS Faint Source Survey.
For comparsion, we have also observed 6 $\lambda$~Bootis stars
where we expect to see typical accretion patterns in the atmospheres.
Some of the objects that 
are accessible from ESO/Chile have been observed before in 
\ion{Ca}{ii}\,K (Lemke \cite{Lem89}, St\"{u}renburg \cite{S93}, 
Holweger \& Rentzsch-Holm \cite{HRH95}, HHK99, but
nothing is known about their surface composition except the Ca abundance. 
We have added one star from the literature to our sample, namely $\beta$~Pic. 
Further details 
on the program stars and the spectral regions observed are 
given in Table~\ref{stars}.

\section{Observations}

We obtained high-resolution spectra with the 1.52m telescope at the 
Observatoire de Haute-Provence equipped with the Aur\'{e}lie 
spectrograph (Gillet et al. \cite{Gil94}) on 4 nights (2000 January 18 -- 
January 22). In the first two nights, we observed the \ion{Ca}{ii}\,K line, 
using grating Nr.\,6, centered at 3934\,\AA , at a 
resolution of R=110\,000. In addition to studying the 
presence of narrow absorption features in the \ion{Ca}{ii}\,K~line, it can also
be used to make a reliable determination of the projected 
rotational velocities. An exact knowledge of this parameter is 
indispensable in our abundance analysis, especially in spectral 
regions with many blend lines and for fast rotators. Furthermore, 
Ca has proven to be a sensitive indicator of diffusion 
processes. We observed Ba for the same reason. 
Sodium (\ion{Na}{i}\,D1 and D2) observations were carried out for the two stars where
we found narrow absorption features in \ion{Ca}{ii}\,K during the
first two nights. In addition, we observed two comparison stars
with pure stellar profiles to correct for telluric lines.
On the third and fourth night we observed \ion{Ba}{ii} and
the \ion{O}{i} triplet at 7771-4\,\AA , using grating Nr.\,5, 
centered at $\lambda_{c}$\,=\,4934\,\AA , and $\lambda_{c}$\,=\,7800\,\AA ,
with R=60\,000. A target list, including the elements observed, is 
given in Table~\ref{stars}.

For eight stars (HR\,804, HR\,1483, HR\,1666, 
HR\,2491, HR\,2763, HR\,5531, HR\,6378, and HR\,8728) we have spectra 
of the \ion{Ca}{ii}\,K line region taken in 1996 and 1997 with the ESO 
CES system, at a resolution of R=70\,000 (see HHK99).

\section{Data reduction}

Data reduction was carried out using ESO MIDAS standard routines. We used 
Th--Ar spectra for the wavelength calibration of the \ion{Ca}{ii}\,K and 
\ion{Ba}{ii} lines, and Th--Ne spectra for the \ion{O}{i} data (Odorico 
et al. \cite{OGP87}).

We have constructed a template to correct for telluric lines in the \ion{Na}{i}\,D 
region by dividing the pure stellar spectrum of HR\,4534
by its synthetic spectrum. This template contains only the telluric
lines, and was used to clean up the spectra of the two stars HR\,1989 and
HR\,6556.

\section{Fundamental parameters}
\label{fundamental}

We obtained values of T$_{\rm eff}$ and $\log g$
from uvby$\beta$ photometry, using an updated version of the UVBYBETA
code (Moon \& Dworetsky \cite{Moon}; Napiwotzki et al. \cite{Napiwotzki}). 
Typical errors for these parameters were $\Delta$~T$_{\rm eff}=200$~K and 
$\Delta \log g=0.1$~dex. The final parameters are given in Table~\ref{tabfund},
columns 3 and 4. Using these parameters and a solar composition 
(Anders \& Grevesse \cite{Anders}), we used the \mbox{ATLAS9} code 
(Kurucz \cite{Kurucz1}) to calculate plane-parallel
atmospheric models. For all stars, we assume a microturbulence, $\xi$, of 
$3.0$~km~s$^{-1}$, consistent with previous work (Holweger \&
Rentzsch-Holm \cite{HRH95}; HHK99).

\begin{table}[h]
\caption{Fundamental parameters of the program stars. The rotational
         velocities denote the $v\sin i$ deduced from the Ca, O and Ba
         spectral range, respectively.}
\begin{tabular}{rrrrc}
\hline\\
 HR    &   HD   &   T$_{\rm eff}$   &  $\log g$   &  $v \sin i$  \\[2mm]
\hline\\
  553  &  11636 &    8370  & 4.1 &  60/75/75 \\
  804  &  16970 &    9230  & 4.1 & 165/175/175 \\
 1483  &  29573 &    8930  & 3.9 &  25/30/30 \\
 1666  &  33111 &    8100  & 3.6 & 180/210/210 \\
 1989  &  38545 &    8600  & 3.5 & 200/190/-- \\
 2491  &  48915 &   10130  & 4.3 &  18/18/-- \\
 2763  &  56537 &    8480  & 3.9 & 150/160/150 \\
 3083  &  64491 &    7140  & 4.1 &  60/25/-- \\
 3569  &  76644 &    8060  & 4.2 & 160/145/145 \\
 4295  &  95418 &    9600  & 3.8 &  43/45/45 \\
 4534  & 102647 &    8630  & 4.2 & 110/125/115 \\
 4828  & 110411 &    9210  & 4.2 &  --/166/-- \\
 5351  & 125162 &    8925  & 4.1 & 105/105/105 \\
 5531  & 130841 &    8240  & 4.0 &  80/80/65 \\
 5793  & 139006 &    9740  & 3.9 & 140/125/115 \\
 5895  & 141851 &    8770  & 3.8 & 250/250/-- \\
 6378  & 155125 &    8850  & 3.9 &  19/15/-- \\
 6556  & 159561 &    7960  & 3.6 & 210/225/225 \\
 7001  & 172167 &    9500  & 3.9 &  22/22/-- \\
 8728  & 216956 &    8760  & 4.2 &  75/--/--\\
 8947  & 221756 &    8800  & 3.8 & 100/100/-- \\
\hline\\
\end{tabular}
\label{tabfund}
\end{table}

\section{Abundance analysis}

Abundance analysis is done by synthesizing a spectrum using
the Kiel line formation code, LINFOR. 
We derive abundances for five elements, O, Ca, Ba, 
Fe, and Y, in the three spectral windows. Here, we describe 
the input line data and the non-LTE calculations for \ion{O}{i}.

  \subsection{Line data}

The line data used in this analysis is taken from the VALD database
(\cite{Kupka}) and is summarized in 
Table~\ref{tabline}. We use a total of 176 lines to fit the spectra
in the spectral range of the Ba and Y lines,
but in most cases, due to heavy blending, the Fe abundance is derived
from a strong blend of Fe lines at 4958\,\AA . We then use this Fe abundance
to derive a Ba abundance from the Ba/Fe blend at 4934\,\AA , and
derive the Y abundance from the Y/Ba blend at 4900\,\AA .

\begin{table}[h]
\caption{Line data used in the abundance analysis: \ion{Y}{ii}, \ion{Ca}{ii} 
and \ion{O}{i} broadening parameters are calculated using the classical 
approximation for 
radiative damping, Griem (\cite{Griem}) and Cowley (\cite{Cowley}) for Stark
broadening and Uns\"{o}ld (\cite{Unsoeld}) for van der Waals broadening. The
last column indicates the reference for all other data. If there are three references
in the last column, the first one refers to the $\log gf$ value, the second one 
to the $\log C_4$ and $\log \gamma$, and the last one to the van der Waals
broadening parameter.}
\begin{tabular}{lllllll}
\hline\\
$\lambda$ [\AA ] &  $\chi_i$     &  $\log gf$  & $\log C_4$ & $\log C_6$ & $\log \gamma$ & Ref. \\[2mm]
\hline\\
\ion{O}{i}       &               &             &            &            &               & \\[1mm]
  7771.94        &\,\,\,9.15     &\,\,0.324    &  -13.25    &   -30.73   &         0.37  & 1 \\
  7774.17        &\,\,\,9.15     &\,\,0.174    &  -13.25    &   -30.73   &         0.37  & 1 \\
  7775.39        &\,\,\,9.15     &   -0.046    &  -13.25    &   -30.73   &         0.37  & 1 \\[2mm]
\ion{Ca}{ii}     &               &             &            &            &               & \\[1mm]
  3933.66        &\,\,\,0.00     &\,\,0.14     &  -13.8     &   -30.95   &         1.34  & 2 \\[2mm]
\ion{Ba}{ii}     &               &             &            &            &               & \\[1mm]
  4899.93        &\,\,\,2.72     &   -0.080    &  -12.57    &   -30.65   &         0.92  & 3 \\
  4934.08        &\,\,\,0.00     &   -0.150    &  -13.17    &   -30.64   &         0.91  & 3 \\[2mm]
\ion{Fe}{i}      &               &             &            &            &               & \\[1mm]
  4933.290       &\,\,\,3.30     &   -2.287    &  -14.82    &   -31.02   &         1.06  & 4 \\
  4933.330       &\,\,\,4.23     &   -0.604    &  -13.61    &   -30.93   &         2.54  & 4 \\
  4934.010       &\,\,\,4.15     &   -0.589    &  -12.98    &   -30.94   &         2.54  & 4 \\
  4934.080       &\,\,\,3.30     &   -2.307    &  -14.81    &   -31.10   &         0.55  & 4 \\
  4957.300       &\,\,\,2.85     &   -0.408    &  -13.71    &   -29.69   &         1.02  & 5/4/6 \\
  4957.600       &\,\,\,2.81     &\,\,0.233    &  -13.71    &   -29.71   &         1.02  & 5/4/6 \\
  4957.680       &\,\,\,4.19     &   -0.400    &  -12.84    &   -30.96   &         2.47  & 4 \\[2mm]
\ion{Fe}{ii}     &               &             &            &            &               & \\[1mm]
  4958.820       &     10.38     &   -0.645    &  -13.52    &   -30.87   &         1.02  & 4 \\[2mm]
\ion{Y}{ii}      &               &             &            &            &               & \\[1mm]
  4900.120       &\,\,\,1.03     &   -0.090    &  -13.36    &   -31.17   &         0.92  & 3 \\
\hline\\
\end{tabular}

1: Wiese et al. (\cite{Wiese}); 2: Wiese et al. (\cite{WSM69})  3: Kurucz (\cite{Kurucz2});\\
4: Kurucz (\cite{Kurucz3}); 5: VALD-2 (Kupka et al. \cite{Kupka});\\
6: Barklem et al. (\cite{Barklem})
\label{tabline}
\end{table}

  \subsection{Non-LTE effects}

We discuss here the non-LTE effects of the individual elements, although
a detailed non-LTE abundance determination was only performed for the O
triplet, where we expect large departures from LTE (Paunzen et al. \cite{Paunzen}).
We derived level populations for all the relevant energy levels of neutral
O, using the Kiel NLTE code, the model atmospheres described 
in Sect.~\ref{fundamental}, and the \ion{O}{i} model atom described in 
Paunzen et al. (\cite{Paunzen}). Non-LTE corrections were typically of 
the order of $-0.5$~dex (see Table~\ref{tababus}).

\begin{table}[th]
\caption{O, Ca, Ba, Y, and Fe abundances from our spectra and from the
         literature (see Sect.~\ref{abus} for details).}
\begin{tabular}{rrrrrrrr}
\hline\\
       &        &  [O]  &  [O]  &  [Ca]  &  [Ba]  &  [Y]  &  [Fe] \\
  HR   &   HD   &  {\tiny LTE}  & {\tiny NLTE}  &        &        &       &       \\[2mm]
\hline\\
  553  &  11636 &  0.4\,\,\,& -0.05     & -0.14  &  1.6\,\,\, &  0.3\,\,\,&  0.0\,\,\,\\
  804  &  16970 &  0.25     & -0.2\,\,\,&  0.16  &  0.2\,\,\, &  0.0\,\,\,&  0.0\,\,\,\\
 1483  &  29573 &  0.0\,\,\,& -0.35     & -0.34  &  1.0\,\,\, &  0.7\,\,\,&  0.0\,\,\,\\
 1666  &  33111 &  0.6\,\,\,&  0.05     & -0.24  & -0.4\,\,\, &  0.0\,\,\,& -0.4\,\,\,\\
 1989  &  38545 &  0.6\,\,\,&  0.0\,\,\,& -0.09  &            &           &           \\
 2020  &  39060 &           &           &  0.02  &  0.09      &           &  0.13     \\
 2491  &  48915 &  0.13     & -0.32     & -0.39  &  1.39      &           &  0.2\,\,\,\\
 2763  &  56537 &  0.65     &  0.10     &  0.00  &  0.2\,\,\, &  0.0\,\,\,& -0.3\,\,\,\\
 3083  &  64491 & -0.62     & -0.8\,\,\,& -1.35  &            &           &           \\
 3569  &  76644 &  0.25     & -0.15     &  0.25  &  0.4\,\,\, &  0.2\,\,\,& -0.1\,\,\,\\
 4295  &  95418 &  0.3\,\,\,& -0.2\,\,\,& -0.16  &  1.0\,\,\, &  0.7\,\,\,& -0.05     \\
 4534  & 102647 &  0.55     &  0.1\,\,\,& -0.04  &  0.3\,\,\, &  0.2\,\,\,& -0.3\,\,\,\\
 4828  & 110411 &  0.4\,\,\,& -0.05     &        &            &           &           \\
 5351  & 125162 &           &           & -1.77  & -0.7\,\,\, &           &           \\
 5531  & 130841 & -0.45     & -0.68     & -0.84  &  0.5\,\,\, &  0.4\,\,\,& -0.7\,\,\,\\
 5793  & 139006 &  0.35     & -0.2\,\,\,&  0.00  &            &           & -0.5\,\,\,\\
 5895  & 141851 &           &           &  0.00  &            &           &           \\
 6378  & 155125 & -0.5\,\,\,& -0.78     & -0.24  &            &  0.26     & -0.51     \\
 6556  & 159561 &  0.6\,\,\,&  0.05     & -0.14  &  0.3\,\,\, &  0.0\,\,\,& -0.4\,\,\,\\
 7001  & 172167 &  0.98     &  0.17     & -0.53  &  1.92      &           & -0.55     \\
 8728  & 216956 &           &           &  0.07  &            &           & -0.03     \\
 8947  & 221756 &           &           & -0.23  &            &           &           \\
\hline\\
\end{tabular}
\label{tababus}
\end{table}

For Fe non-LTE corrections have been shown to be well below 
$+0.25$~dex for all program stars (Rentzsch-Holm \cite{Rentzsch}).
We did not perform NLTE calculations in this case since the typical 
error from our abundance analysis was around 0.2~dex.

Non-LTE corrections of Ba have been shown to depend strongly on 
the Ba abundance itself (Lemke \cite{Lemke2}). In Lemke's sample 
of normal A stars, the abundance corrections range from $-0.07$ to $+0.30$~dex. 
The overabundances we derived in this work are up to 2.0~dex, so can never
be removed by non-LTE corrections.

Since we observe \ion{Y}{ii} and \ion{Ca}{ii}, we do not expect significant 
departures from LTE, as these are the dominant ionization states of the atom. 
This has been justified by St\"{u}renburg (\cite{S93}), who found a mean 
correction of 0.03~dex for the \ion{Ca}{ii}\,K
resonance line.

  \subsection{Abundances}
  \label{abus}

The spectral abundances derived from this work are summarized in 
Table~\ref{tababus}. It proved impossible to derive reliable
Ba and Y abundances in the star HR\,5793 due to its large rotational 
velocity and the poor S/N ratio in the spectral range of the Ba lines.
Given the large number of Fe lines we were able to
derive an Fe abundance. In HR\,5351, we could only deduce
the Ba abundance. The typical error in the 
abundance analysis is 0.02--0.05~dex, depending
on the spectral quality. In addition, the uncertainty in the stellar 
parameters may introduce an error of the order of 0.2~dex.

In Vega, our Ca abundance, $\log \epsilon({\rm Ca})=5.83$, compares 
well with the value of 5.82, obtained by Lemke (\cite{Lemke2}).
The non-LTE O abundance, [O]$=0.17$, derived from the triplet at 7771~\AA ,
also agrees well with the LTE abundance derived from 
visible lines that are barely affected by non-LTE, [O]$=0.18$, 
(Qiu et al. \cite{Qiu}).
In Sirius, our \ion{Ca}{ii} abundance of 5.97 could indicate that
Lemke (\cite{Lemke2}) slightly overestimated the non-LTE effects in
neutral Ca when deriving Ca abundances, $\log \epsilon({\rm Ca,LTE})=5.65$,
$\log \epsilon({\rm Ca,NLTE})=6.26$.

We completed our data using the following abundance analyses;
Lemke (\cite{Lem89}, \cite{Lemke2}) for HR\,2491 (Ba, Fe): 
Gigas (\cite{Gigas1}, \cite{Gigas2}) for HR\,7001 (Ba, Fe):
Holweger et al. (\cite{Holweger97}) for HR\,2020 (Ca, Ba, Fe): Dunkin et al. 
(\cite{Dunkin}) for HR\,8728 (Fe): Cowley \& Aikman (\cite{Cowley1}) 
for HR\,6378 (Y, Fe).

\section{Circumstellar lines}

Our analysis of the \ion{Ca}{ii}~K data gave evidence for
narrow absorption features in the rotationally broadened stellar line profile
of two stars, HR\,1989 and HR\,6556.

\begin{figure}[h]
\vspace*{2mm}
\resizebox{\hsize}{!}{\includegraphics{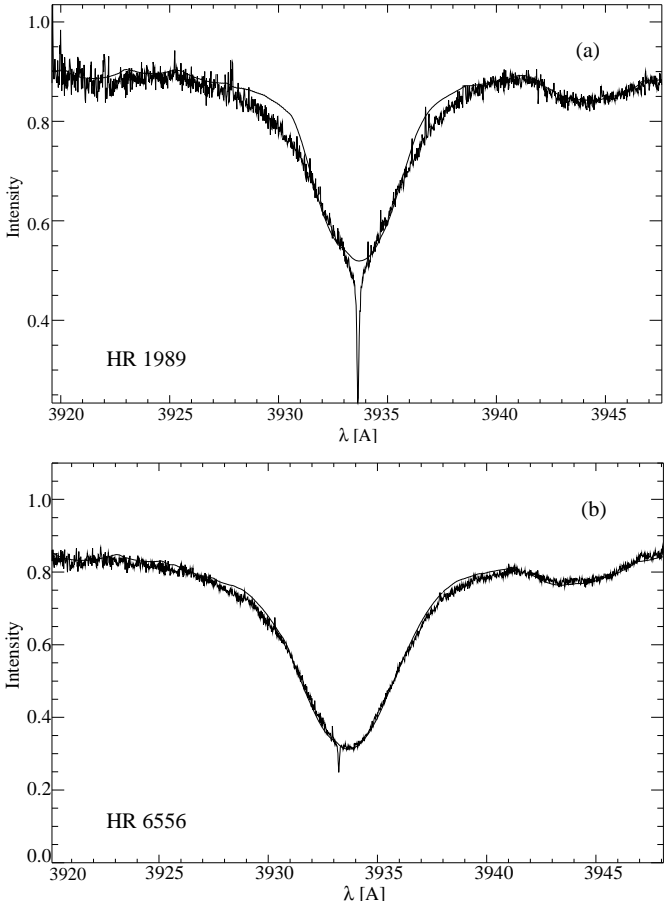}}
\caption{High resolution OHP spectra of the \ion{Ca}{ii}~K line region:
a) HR\,1989, b) HR\,6556. Synthetic spectra using the parameters of
Tables~\ref{tabfund} and \ref{tababus} are overplotted.}
\label{figcs}
\end{figure}

\subsection{HR\,1989}

Narrow absorption lines had already been detected in HR\,1989 by St\"{u}renburg 
(\cite{S93}), Bohlender \& Walker (\cite{Bohlender}), and
Hauck et al. (\cite{Hauck}). The latter gives an equivalent
width of 72.2~m\AA , at a redshift of 11.1~km~s$^{-1}$. Our
data gives a similar equivalent width, of 65~m\AA , but at a 
blueshift of 1~km~s$^{-1}$ (Fig.~\ref{figcs}a). Therefore we conclude that 
the absorption feature varies with radial velocity with respect to the star, 
proving that in this case it is of circumstellar origin.
 
Despite the poor S/N ratio of the spectrum, the \ion{Na}{i}\,D lines 
in HR\,1989 show strong narrow absorptions at the line centers.
Their equivalent width are 72.2 and 33~m\AA\  for the 
D2 and D1 line respectively, but the uncertainty on these values is large
since the stellar line fit is so poor. Sfeir et al. (\cite{Sfeir}) obtained 
data for lines-of-sight towards 143 stars, one being HR\,1989, to map the local 
bubble. They derived equivalent widths of 34.1 and 16.4~m\AA\  for the \ion{Na}{i}\,D2 
and D1 lines, respectively. Using our Ca observations and Sfeir et al. (\cite{Sfeir})
Na observations, we obtained a \ion{Na}{i}/\ion{Ca}{ii} ratio that is lower than 1.
This also hints at a circumstellar rather than interstellar origin of the
features provided that we ignore possible variabilities between the different 
observing periods.

\subsection{HR\,6556}

In HR\,6556, HHK99 found a circumstellar Ca absorption with an 
equivalent width of 22~m\AA ,  at a blueshift of 35~km~s$^{-1}$.
Our data gives a slightly smaller equivalent width of 14~m\AA , at a 
blueshift of 32~km~s$^{-1}$ (Fig.~\ref{figcs}b). Therefore this feature
appears to be stable within the error limits.
The Na spectrum of HR\,6556 is so noisy that we are not able to 
deduce the presence or absence of narrow circumstellar lines.

A study of the interstellar \ion{Na}{i} density distribution in the solar neighbourhood
revealed that the observed Na column density in HR\,6556 did not
agree with neighbouring values (Vergely et al. \cite{Vergely}). In
fact, the discrepancy is about a factor of 200. We interpret this as
evidence for circumstellar gas around HR\,6556.

\section{Dusty versus dust-free A stars}

To ensure that our sample was homogeneous, we removed all the newly discovered 
spectroscopic binaries (see Table~\ref{stars}), 
since the abundance determination in these objects underestimates
the real abundances. We also excluded the $\lambda$~Bootis stars, because
these belong to a subgroup of A stars where the mechanism giving rise to the
metal-poor abundance pattern is not yet fully understood (Paunzen \cite{P99}). 
In addition some of the $\lambda$~Bootis stars are beyond 50~pc, and IRAS
measurements give only upper limits on the infrared fluxes, so that we do 
not know whether they are
dusty or not. The remaining stars were split into two groups, those with
(5) and without (7) infrared excess.

\begin{figure}[ht]
\vspace*{-3mm}
\resizebox{\hsize}{!}{\includegraphics{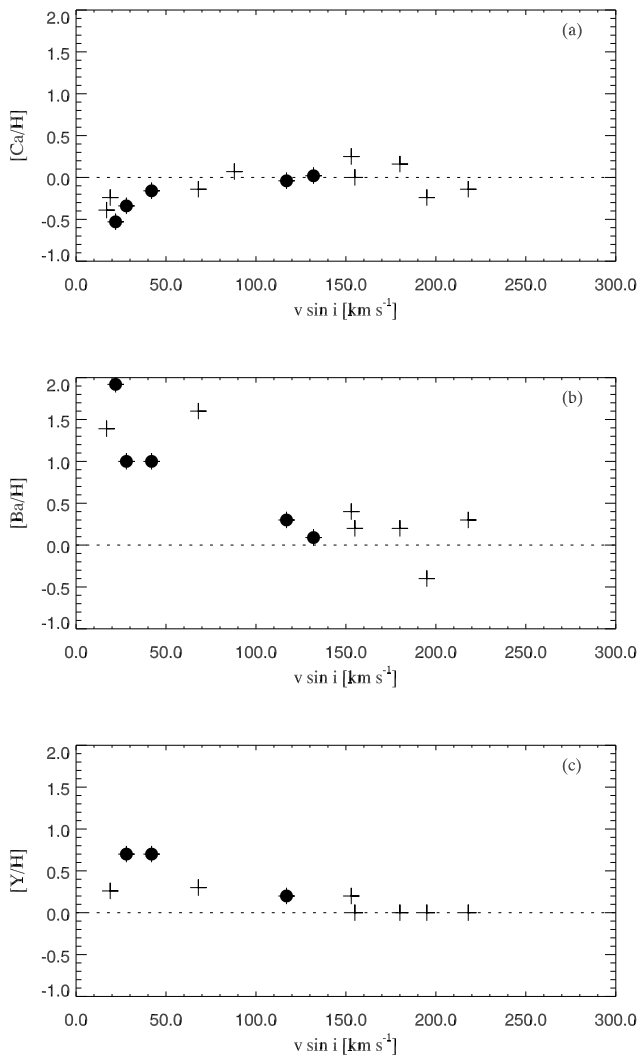}}
\caption{Photospheric Ca, Ba and Y abundances versus stellar rotation:
         dusty A stars (plus signs with filled circles) and dust-free A stars (plus
         signs)}
\label{figvsini}
\end{figure}

Even small amounts of accreted material will show up in the photospheric
composition of A stars, since they have shallow convection zones 
($\sim 10^{-9}$~M$_\odot$, Turcotte \& Charbonneau \cite{Turcotte}).
Besides accretion, meridional mixing and 
diffusion are the most important processes affecting the stellar 
abundance pattern. While meridional mixing simply wipes out any
inhomogeneities, diffusion acts selectively on certain elements,
imprinting its own abundance pattern which differs from the 
accretion pattern.

Surprisingly, there is no difference between the abundance patterns 
of dusty A stars and those of dust-free A stars. Table~\ref{statistik}
shows that the mean abundance patterns of both groups of stars agree
within the abundance analysis error. Only for Y do we have a 
discrepancy, which can be explained by the fact that this element is only 
determined in very few stars.

\begin{table}[h]
\caption{Mean abundance pattern for the dusty (first row) and dust-free 
(second row) A stars; in
parentheses we give the standard deviation (in units of 0.01~dex)}
\begin{tabular}{rrrrr}
\hline\\
 {\mbox [O]}\hspace*{5mm}& [Ca]\hspace*{5mm}& [Ba]\hspace*{5mm}& [Y]\hspace*{5mm}& [Fe]\hspace*{5mm}\\[2mm]
\hline\\
 -0.15 (19) &  -0.09 (15) &   0.60 (41) &  0.53 (24) & -0.05 (14) \\
 -0.07 (14) &  -0.07 (21) &   0.53 (66) &  0.08 (12) & -0.14 (21) \\
\hline
\end{tabular}
\label{statistik}
\end{table}

We do not find any correlation between abundances and effective 
temperature or gravity.
However there is a pronounced correlation between the Ba
abundances and the rotational velocity, with the overabundances
diminishing with increasing $v \sin i$ (Fig.~\ref{figvsini}b). According to 
Michaud (\cite{Mich70}), the overabundances of Ba are due to diffusion 
processes, and their disappearence at higher $v \sin i$ is a result of efficient 
meridional mixing at rotational velocities in excess of 100~km~s$^{-1}$
(Turcotte \& Charbonneau \cite{Turcotte}). The Y abundances show also a 
marginal increase towards lower $v \sin i$. Cowley (\cite{Cowley0}) noted that
diffusion theory does not predict large Y overabundances, which is in 
good agreement with our findings. We expect mild 
underabundances of Ca from diffusion theory, a trend which is qualitatively 
seen for stars with low rotational velocities in Fig.~\ref{figvsini}a. In all 
three cases, no distinction can be made between the dusty and dust-free
stars in our sample.

\begin{figure*}[ht]
\resizebox{\hsize}{!}{\includegraphics{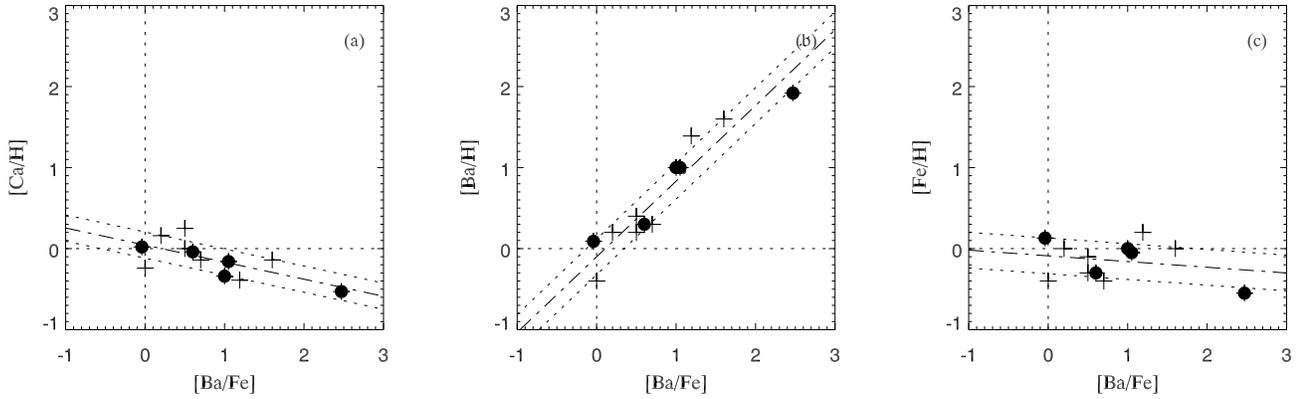}}
\caption{Stellar abundances versus the diffusion indicator [Ba/Fe]:
         dusty A stars (plus signs with filled circles) and 
         dust-free A stars (plus signs). The dash-dotted lines
         gives the straight line fit to the whole data set
         with the rms deviation indicated by the dotted parallel lines.}
\label{figdiff}
\end{figure*}

A plot of the specific element abundance versus [Ba/Fe] ratio is used
as a diagnostic tool to detect diffusion signatures. The use of the ratio
ensures that any
star-to-star metallicity variation cancels, and the influence
of accretion likewise is excluded --- in the standard accretion model Fe and
Ba are both entirely condensed in dust grains --- so that we are
left only with diffusion effects. Fig.~\ref{figdiff}a and \ref{figdiff}b 
illustrate how Ca and Ba are affected by diffusion: Ba shows large
overabundances, but there is a trend for Ca to be depleted in the 
photosphere in the
presence of strong diffusion. This is in perfect qualitative 
agreement with the theoretical predictions of Michaud (\cite{Mich70}).
As expected, we find that in our program stars, O and Fe are not
affected by diffusion due to their high abundance and large number of lines
(Fig.~\ref{figdiff}c).

We derive least-square fits for the correlations between the Ca, Ba and Fe 
abundances with the [Ba/Fe] ratio, using the dusty and dust-free
stars (dashed-dotted lines in Fig.~\ref{figdiff}). The parameters of 
the individual fits are calculated assuming a typical abundance
error of 0.2~dex for all stars. The results are summarized in 
Table~\ref{statistik1}. Based on this analysis, we can 
definitely exclude a pure accretion signature for the dusty 
A stars.

\begin{table}[h]
\caption{Statistical parameters for the sample of dusty and dust-free
A stars: the best fit is given by [X]$=$m$\cdot$[Ba/Fe]$+$c for the 
combination of both samples with the corresponding rms value and
the rms values of the individual samples (1 - dusty, 2 - dust-free).}
\begin{tabular}{lrrr}
\hline\\
        &   [Ca]\,vs.\,[Ba/Fe]  &  [Ba]\,vs.\,[Ba/Fe] & [Fe]\,vs.\,[Ba/Fe] \\[2mm]
\hline\\
m       & -0.21$\pm$0.08 &  0.93$\pm$0.08 & -0.07$\pm$0.08 \\
c       &  0.04$\pm$0.14 & -0.10$\pm$0.14 & -0.09$\pm$0.14 \\
rms     &  0.16\hspace{8.3mm} &  0.22\hspace{8.3mm} &  0.22\hspace{8.3mm} \\
rms$_1$ &  0.06\hspace{8.3mm} &  0.19\hspace{8.3mm} &  0.19\hspace{8.3mm} \\
rms$_2$ &  0.17\hspace{8.3mm} &  0.21\hspace{8.3mm} &  0.21\hspace{8.3mm} \\
\hline
\end{tabular}
\label{statistik1}
\end{table}

\section{Discussion}

We showed that dusty and dust-free A stars possess the same abundance patterns and
fit both into the diffusion scenario. We do not detect any specific 
accretion signature in the stars with circumstellar dust, even though there are a few
slow rotators amongst them. This result agrees with the findings
of Holweger et al. (\cite{Holweger97}) for $\beta$~Pictoris, which, despite of its
conspicuous circumstellar disk, shows no signs of accretion. 
Typical disk masses derived for the dusty stars of our sample range from
$10^{-8}$ to $10^{-4}$~M$_\odot$. So in all cases the disk contains enough material
to contaminate the stellar photosphere. In their study Cheng et al. (\cite{Cheng92}) 
excluded any confusion of the infrared-excess with a background source, 
which leaves us with a question: Why do the dusty A stars seem 
to be unaffected by the surrounding circumstellar dust?

There are two possibilities which may explain the non-detection
of any peculiar abundance pattern in dusty A stars. First, the stars may accrete 
circumstellar matter at an extremely low rate, e.g. 
$\stackrel{\cdot}{M} < 10^{-14}$~M$_{\sun}$~yr$^{-1}$. This would mean 
that the accretion rate is too low to overcome diffusion and therefore
cannot significantly contaminate the convection zone of the star. The second 
possibility is that there is no interaction between the star 
and the disk, implying the existence of a gap, similar to that which 
is apparant in the case of $\beta$~Pictoris and HR\,4796.

\begin{acknowledgements}
We thank the DFG for subsidizing this project by a travel and observing grant
(KA 1581/1-1) and Helen~Fraser for a careful reading of the manuscript. 
\end{acknowledgements}

\end{document}